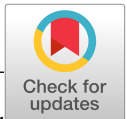

## SPACE SCIENCES

# Formation of extraterrestrial peptides and their derivatives

Serge A. Krasnokutski[1]*, Cornelia Jäger[1], Thomas Henning[2], Claude Geffroy[3], Quentin B. Remaury[3], Pauline Poinot[3]*

The formation of protein precursors, due to the condensation of atomic carbon under the low-temperature conditions of the molecular phases of the interstellar medium, opens alternative pathways for the origin of life. We perform peptide synthesis under conditions prevailing in space and provide a comprehensive analytic characterization of its products. The application of $^{13}C$ allowed us to confirm the suggested pathway of peptide formation that proceeds due to the polymerization of aminoketene molecules that are formed in the $C + CO + NH_3$ reaction. Here, we address the question of how the efficiency of peptide production is modified by the presence of water molecules. We demonstrate that although water slightly reduces the efficiency of polymerization of aminoketene, it does not prevent the formation of peptides.

## INTRODUCTION

One of the most widely accepted scenarios for the origin of life on Earth assumes a spontaneous transition from nonliving to living matter under certain conditions, for example, in a primordial soup of organic chemicals that existed on early Earth (1). It is assumed that over time, these prebiotic molecules started to form more complex molecules that eventually led to the formation of biopolymers with catalytic functions, such as peptides and RNA. These biocatalysts may facilitate the further formation of biomolecules and lastly lead to the autocatalytic chemistry that would start natural selection (2–4). This point can be considered the beginning of the emergent properties of living systems.

Although some of the organic molecules necessary for the origin of life can also be formed on Earth [e.g., (5, 6)], delivery from space could provide distinct advantages because it can deliver these molecules in concentrated form (7). In addition, delivery from space can provide a wider range of organic molecules than might be possible through terrestrial processes alone. Some organic molecules may be formed in space environments that were not present on Earth before the emergence of life or may be formed more efficiently due to the unique conditions found there. Delivered to different locations on Earth that are shielded from the destructive effects of ultraviolet radiation, such as small lakes, ponds, or the cracks or crevasses in rocks, these organic compounds could have persisted and accumulated over time, increasing the likelihood that they will eventually play a role into the synthesis of biomolecules.

Several types of biomolecules such as amino acids, nucleobases, sugars, phosphates, and lipids that are important for different biological functions can be formed in space. These molecules were detected in meteorites and comets (8–10). They could reach Earth and exoplanets through a variety of processes, such as the accretion of comets and cometary-type pebbles, meteorites, interplanetary dust particles, and impacts of asteroids. In particular, the massive delivery of biomolecules during the period of heavy bombardment about four billion years ago has been proposed to play a role in the origin of life (7, 11).

Recently, peptides were shown to form efficiently under the low-temperature conditions typically found in the interstellar medium (ISM) (12). It was suggested that synthesis occurs on the surface of dust particles present in translucent molecular clouds. This material can then be incorporated into the denser phases of molecular clouds and circumstellar disks during the star and planet formation process. Last, these dust grains are building blocks of comets and asteroids. Comets are expected to have the least processed material, and they can drift radially inward in disks approaching habitable zones (13). The asteroids and meteorites should contain more processed material. However, such processing will not destroy all biomolecules, and peptides were also recently potentially detected in meteorites (14).

One of the key properties of peptides is their ability to catalyze a wide range of chemical reactions, including the formation of peptide bonds themselves (15, 16). Peptides with only five amino acids in the chain can already exhibit the appearance of a secondary structure, which is often linked to enhanced catalytic activity (17). Peptides may have also played a role in the formation of the first cell membranes (18, 19). Therefore, the major role of peptides in the origin of life was suggested (1, 20, 21). On early Earth, peptide synthesis typically considers occurring through the polymerization of amino acids. This process involves breaking chemical bonds leading to detaching water molecules, which creates a high energy barrier for polymerization. Furthermore, the reverse process of hydrolysis can cause forming peptides to break down. Therefore, the formation of peptides was considered to occur via the action of catalysts or different concentration processes, such as wet-dry cycling (22). The delivery of ready-made peptide chains to early Earth could be an important alternative or complementary pathway toward the origin of life. It is now very challenging to determine the significance of extraterrestrial delivery compared to the potential route of biopolymer formation directly on Earth.

The recent discovery of a peptide formation pathway operating under ISM conditions with CO, C, and $NH_3$ as reactants indicates that peptides were indeed most likely delivered on early Earth (12). A comprehensive analytical characterization of the molecules formed in such a condensation process is highly desirable.

[1]Laboratory Astrophysics Group of the Max Planck Institute for Astronomy at the Friedrich Schiller University Jena, Helmholtzweg 3, D-07743 Jena, Germany. [2]Max Planck Institute for Astronomy, Heidelberg, Germany. [3]Institut de Chimie des Milieux et Materiaux de Poitiers, University of Poitiers, UMR CNRS 7285, France.
*Corresponding author. Email: sergiy.krasnokutskiy@uni-jena.de (S.A.K.); pauline.poinot@univ-poitiers.fr (P.P.)

Moreover, the previous experiments were performed under ideal conditions when only reactants required for the peptide chain formation were present. Peptide formation in natural environments is expected to be the most efficient in translucent molecular clouds and at the surface of planet-forming disks. Translucent molecular clouds are a stage in molecular cloud evolution, where most carbon exists in the form of single neutral carbon atoms, and slow oxidation leading to the conversion of carbon to CO has started. The temperature of the dust particles in these clouds is already very low, and grains are covered by molecular ice, where CO and $NH_3$ molecules are the main components (23). Therefore, the reactants involved in the reaction of peptide formation are among the most common in such molecular clouds, which suggests a high efficiency of peptide formation under these conditions. Surface regions of planet-forming disks may provide similar conditions. However, water is the most abundant molecule in molecular ices. Its impact on the pathway for peptide formation has yet to be investigated.

Water can react with C atoms leading to the formation of $H_2CO$ molecules (24). However, this reaction is rather slow at low temperatures. The reaction proceeds due to tunneling, which is responsible for the transfer of protons from oxygen to carbon. Although water clusters of specific sizes could act as catalysts and lower the barrier for this reaction (25), its rate should still be much lower compared to barrierless reactions such as $C + NH_3 \rightarrow HCNH_2$ (26) and following reaction $HCNH_2 + CO \rightarrow H_2NCHCO$ (12, 27) leading to the formation of aminoketene molecule. This is also in line with the fact that C atoms can diffuse on the surface of water ice (28); therefore, the encounters of C atoms with other molecules are greatly enhanced. The aminoketene molecules efficiently polymerize even at low temperatures leading to the formation of peptides (12). These theoretical expectations about the role of water have yet to be verified experimentally.

## RESULTS

### Analytical characterization of peptides

Figure 1 shows the results of ultrahigh-performance liquid chromatography analysis combined with triple quadrupole mass spectrometry (UPLC-TQ MS) of the room temperature residual (RTR) obtained by deposition of $^{13}C$, CO, and $NH_3$ reactants on the substrates at 10 K and subsequent warm-up to 300 K. For these experiments, $^{13}C$ was used as a reactant instead of $^{12}C$ to eliminate the possibility of natural $^{12}C$-gly peptides detection and prevent data misinterpretation. Preliminary analysis of the RTR by UPLC–high-resolution mass spectrometry (HRMS) confirmed the absence of $^{12}C$-gly derivatives (refer to table S1). Therefore, only $^{13}C$-peptides were monitored further by TQ MS. During the experiments, $^{13}C$ atoms and $^{12}CO$ gas were used, resulting in a $^{13}C/^{12}C$ ratio of approximately 0.5. This ratio eliminates the possibility of these molecules being contaminants and provides additional evidence supporting the proposed chemical pathway of peptide formation. However, to ensure $^{13}C$-gly peptide identification, we used UPLC-TQ MS that specifically detects a given molecule based on the match of its fragmentation pattern and the retention time with those obtained from the analytical standards. Hence, it targets a compound by monitoring simultaneously intact molecules (named the precursor ion) and a particular fragment ion. The pair of a precursor and fragment ion is named "transition" (table S2), thus enabling better discrimination between structural isomers compared to UPLC-HRMS analysis. Following this strategy, we targeted specific transitions of the most abundant peptide series ($NH_2$- $Gly_n$-$NH_2$) found in RTR in our previous study (12) to give evidence for their presence in the RTR.

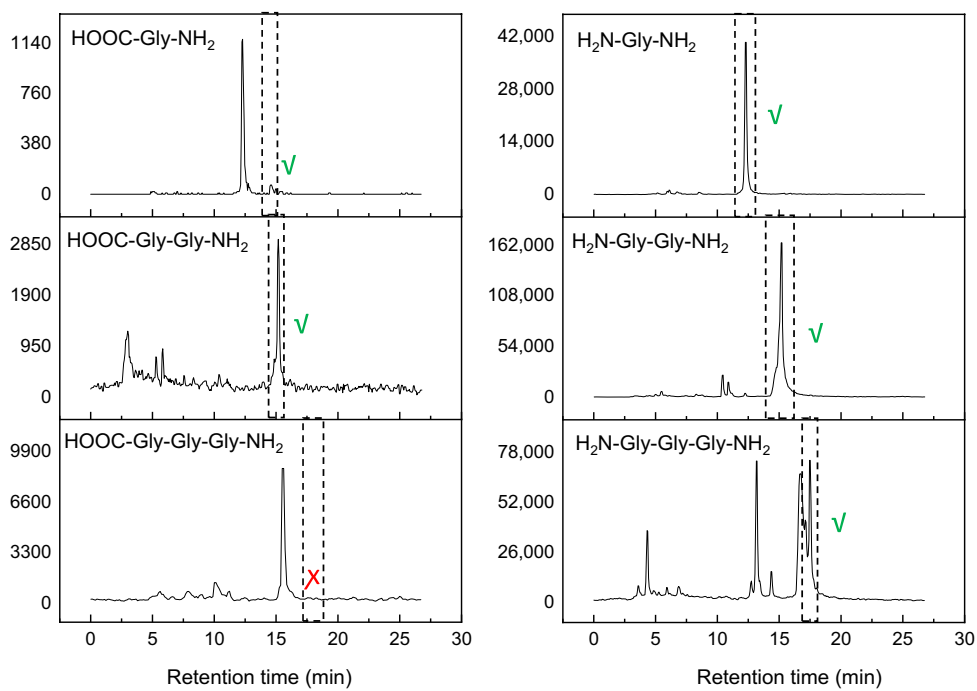

**Fig. 1. UPLC analysis of the extract.** Ion signal on the masses of the specified molecules as a function of the retention time from ultrahigh-performance liquid chromatography analysis of the extract from room temperature residual ($^{13}C$, CO, and $NH_3$ reactants). The appearance times of ions from chemical standards are depicted by the dotted lines. The positive identification of respective molecules is indicated by green checkmarks, while a red cross stands for negative identification.

Figure 1 demonstrates the comparison of the ion signals on the masses $NH_2$-$Gly_n$-$NH_2$ from RTR and the same $^{12}C$ molecule standards purchased commercially, which are shown in fig. S1. As can be seen, there is a good match in the retention times between the standards and the $NH_2$-$Gly_n$-$NH_2$ molecules extracted from RTR. This confirms the previous assignment of the molecular structures of the investigated ions. For the glycine trimers, however, there are several additional peaks at earlier retention times. Ions appearing at these early times have the same fragmentation pattern as ions appearing at expected times. It suggests that the molecular structures of these isomers are very similar to that of linear $NH_2$-$Gly_3$-$NH_2$ peptides. In particular, the ions appearing at very early retention times are very interesting. A higher molecular mobility in chromatographic columns is likely caused by the more hydrophobic properties that nonlinear isomers may have. Since such isomers are intrinsically nonlinear and have a secondary structure, they could potentially mimic the structures of enzyme catalytic centers providing them with catalytic activity. For example, the cyclization of short peptides is often required to achieve high catalytic activity (29–31).

Moreover, this analysis allowed us to detect the canonical glycine and glycine dipeptide with $NH_2$ and COOH terminals. The signal intensity of the molecules of canonical dipeptides was about 60 times lower compared to the same dipeptide with amino terminals. Since the water was not present during the condensation, the chemical pathway of COOH terminal formation is not completely clear and perhaps happened due to the hydrolysis during the solvation of RTR in a water-methanol mixture. The fraction of canonical peptides formed in nature may be greater than those observed in laboratory experiments, as hydrolysis occurs over a much longer period of time. We can exclude the possibility that the detection of canonical peptides was due to the presence of either biological contamination or impurities from previous experiments since we used $^{13}C$ isotopes of C atoms for the formation of RTR. Besides the reliable detection of peptides, the use of $^{13}C$ isotopes also allowed us to confirm the suggested pathways of peptide formation. The formation of peptides was suggested to happen due to the polymerization of aminoketene ($NH_2$CHCO) molecules, which form in the reaction of C + $NH_3$ + CO (12, 27). In this case, the use of $^{13}C$ atoms during the condensation should result in the presence of one $^{13}C$ atom per glycine residual in the peptide chain. As can be seen in Fig. 1, exactly such molecules were formed in our experiments.

To further confirm the formation of peptides in our experiments, we applied a recently developed proteomics-based strategy, which was invented for the investigation of peptide sequences in extraterrestrial samples (14). The main idea of this method is rather simple: Aminopeptidase M (APM) is used to cut peptide bonds from the N-terminal end of peptides. The schematic diagram of APM function is shown in fig. S2. Therefore, if any sample contains peptides, then the incubation of this sample with APM should reduce the number of peptide molecules and, accordingly, lead to an increase in the number of amino acid molecules in this sample. The quantities of peptide and amino acid molecules were monitored using UPLC-TQ MS in the same way that was used to collect the data shown in Fig. 1. Figure 2 shows the summary of these studies. As can be seen, we indeed observe a considerable reduction in the signal intensity of all targeted peptides and an increase in or appearance of glycine ($NH_2$CH$_2$COOH) and glycinamide ($NH_2$CH$_2$CONH$_2$). Therefore, these results additionally support the presence of glycine peptides in the RTR.

## Role of water

Figure 3 shows the infrared (IR) absorption spectra of RTR obtained by deposition of reactants on the substrates at 10 K and subsequent warm-up to 300 K. Both RTRs were produced under the same conditions. The only difference was the presence of water in the gas mixtures used to produce ice. As can be seen, the presence of water has a relatively small effect on the absorption spectra of the RTRs. The main difference is that the peptide absorption bands of the RTR obtained in the presence of water are approximately two times weaker. A reduction of their intensities shows a smaller amount of peptide bonds in the RTR sample when water was added. To better characterize the formed molecules, we used the same ex situ mass spectrometry analysis as was used in our previous studies. The RTR was solvated in a methanol-water mixture and provided to the mass spectrometer equipped with an electrospray ionization (ESI) source. As can be seen in Fig. 4, the formation of various peptides is observed, which is in line with in situ IR spectroscopy measurements. The main difference, however, is that the main products are not $NH_2$-$Gly_n$-$NH_2$ peptides but the more direct polymers of aminoketene with only an extra hydrogen atom. The peptides formed in this experiment have $NH_2$ and COH terminals. Moreover, the balance is shifted toward the formation of short peptides. In this experiment with water, we observe the dominant formation of dimers with a considerably lower number of trimers, while in the experiments without water, the mass peak assigned to glycine trimers with amino terminals on both sides was the most intense one (12). Therefore, we conclude that ammonia, as originally suggested, was indeed very important for the polymerization of aminoketene molecules. When water molecules are present in the ice, the increase of the substrate temperature leads to the evaporation of ammonia before water. Therefore, during ammonia evaporation, aminoketene molecules will be trapped in the water ice matrix without the possibility to move freely. The aminoketene molecules will be able to meet each other only during the evaporation of water when most of the ammonia is already evaporated. This supports the important catalytic role of ammonia in the polymerization of aminoketene and is in line with the previous finding that ammonia catalyzes the polymerization of $H_2CO$ to polyoxymethylene (*32*), which is the first detected polymer in space (*33*). Our findings are also in line with the predicted low rate of the $H_2O + C \rightarrow H_2CO$ reaction (*24*) and efficient diffusion of the C atoms on the surface of water ice (*28*) since the amount of $H_2CO$ molecules produced in current experiments with water was below the reliable detection limit as can be seen in fig. S3.

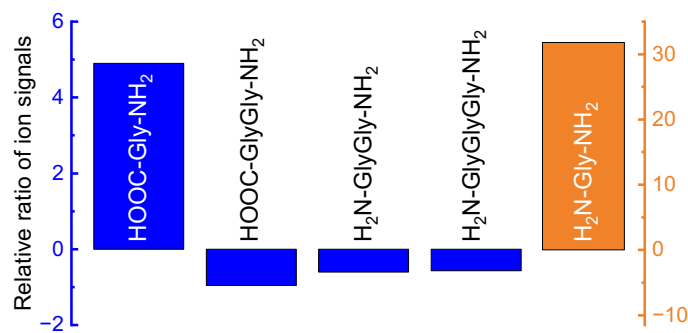

**Fig. 2. The impact of enzyme on the room temperature residual.** Relative ratio of ion signals after incubation of room temperature residual ($^{13}C$, CO, and $NH_3$ reactants) in the absence and in the presence of APM. Relative ratio calculated as (ion signal with APM − ion signal without APM)/ion signal without APM.

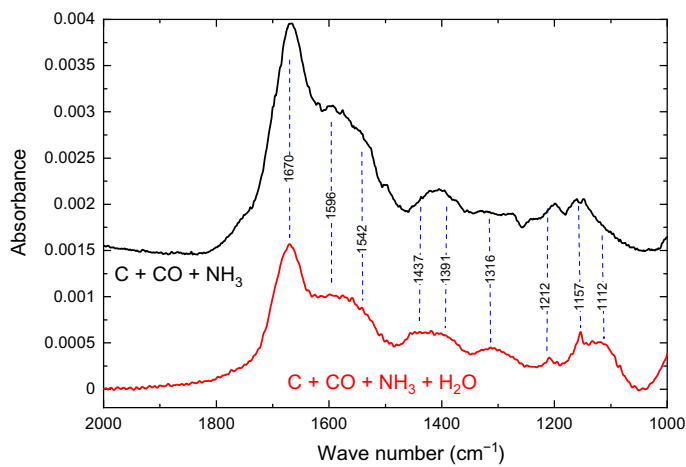

**Fig. 3. The IR absorption spectra of room temperature residuals.** Room temperature residuals are produced under identical conditions, with and without the presence of water. In both experiments, the amount of the limiting reactant, specifically C atoms, was equal. The upper spectrum was shifted by the absorbance scale for the clarity purpose.

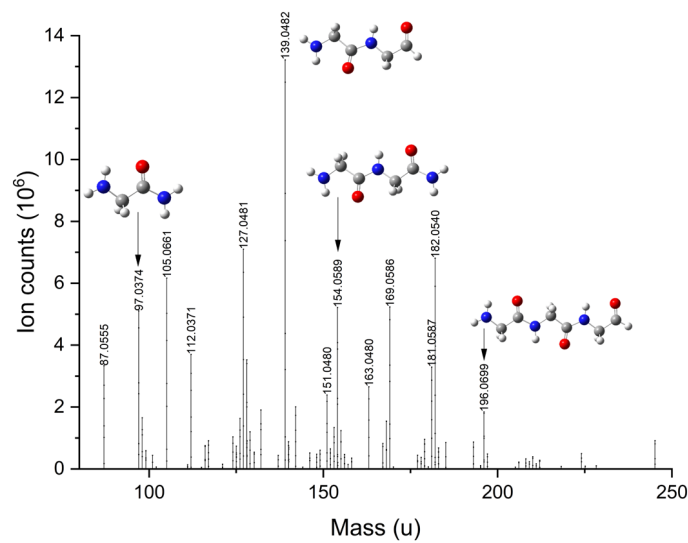

**Fig. 4. Mass spectrum of the soluble extract from room temperature residual ($C$, $CO$, $NH_3$, and $H_2O$ reactants).** The molecular structures assigned to the peaks are shown without $Na^+$, which is added during the ionization.

## DISCUSSION

### Astrophysical implication

Our experiments can be compared to the case of single isolated dust particles that are present in molecular clouds of the ISM. In the denser parts of molecular clouds, dust particles start to coagulate—a process which continues in planet-forming disks until celestial bodies are formed. Comets that are formed in the disks with large separations from new stars are the most interesting objects for the formation of peptides. When the temperature of a comet rises, only the molecules from a thin surface layer can evaporate freely. The evaporation of the molecules from the interior is largely suppressed. When the temperature of such a body rises to about 176 K, ammonia present in the molecular ice will concentrate by transferring into the liquid phase forming a eutectic melt with water ($NH_3 \cdot 2H_2O$) (*34*). Experimental detection of traces of water activity on comets and asteroids supports the presence of liquid phase in these bodies (*35*, *36*). The temperature of comets and asteroids is expected to increase due to their proximity to the sun or the decay of short-lived radioactive isotopes, such as $^{26}Al$ (*37*). These conditions could be nicely suited for the polymerization of the aminoketene molecules that are formed in molecular clouds due to the accretion of C atoms on the surface of dust particles. The liquid phase allows the free movement of the soluble molecules. Such a melt provides a high concentration of ammonia molecules that act as catalysts. Last, the long astronomical timescales ensure enough time for chemistry to take place. The last point can be practically important since we detected only a very small number of RTR as can be seen in fig. S4 when condensation of the same reactants ($C$, $CO$, and $NH_3$) took place inside liquid helium droplets flying in a vacuum chamber. Moreover, no peptides were found in this RTR. Therefore, a very quick warm-up prevents polymerization. The same dependence of polymerization from the time of warming up was also presumably observed in our original experiment with ice deposited on substrates. However, because of the large fluctuation in the sensitivity of the ex situ analysis, it cannot be confirmed unambiguously. The time dependence can be easily understood considering that the chemistry takes place in the solid state when the free movement of the molecules is still restricted, and consequently, the achievement of the right arrangement of the molecules requires time. Moreover, the participation of the catalyst, although reducing the energy barriers for the polymerization reactions, likely does not make them equal to zero. Overcoming the energy barriers or tunneling through these barriers requires time.

### Possible polymerization pathways and wet chemistry

The polymerization process of aminoketene molecules ($NH_2CHCO$) involves transferring a proton from the $NH_2$ group. This results in the formation of an NH group that is present in peptide chains and activates the nitrogen, allowing it to chemically bond with another aminoketene molecule forming a peptide bond HNCO. Proton transfer can occur either intra- or intermolecularly. Figure S5 shows that intramolecular proton transfer has a relatively high barrier. This is also applicable to established peptides. Therefore, low-temperature chemistry should proceed either via tunneling, which is common for protons, or with the assistance of catalyst molecules (*12*). Moreover, intramolecular proton transfer results in the fragmentation of aminoketene with the formation of $H_2CNH$ and CO molecules. $H_2CNH$ can be added to aminoketene molecules resulting in the formation of stripped dipeptide with $NH_2$ and $CH_2$ end groups as shown in the bottom frame of fig. S5. Peptides do not fragment in case of intermolecular proton transfer when H from the $NH_2$ group is transferred to any other molecule, or in the case of a new chemical bond between the C atom of the CO group is established simultaneously with the proton transfer. The peptide in the established form is also not fragmented, as shown in the bottom frame of fig. S5. A better understanding of the polymerization pathway could be obtained from molecular dynamic simulations of large ice clusters. Our experiments strongly support the scenario with intermolecular proton transfer. In the experiments where water was absent, ammonia served as the sole source of hydrogen. Therefore, the efficient intermolecular proton transfer is demonstrated by the identification of numerous peptides in these experiments that exhibit an N/H ratio that differs from that observed in ammonia. As discussed above, the results with water point to $NH_3$ as a catalyst in the polymerization process. The role of ammonia could be to act as a proton donor and acceptor. Proton transfer is a process that generally happens with a high probability. For example, in the case of ion-molecular reactions, proton transfer, when exothermic, is considered to occur at the collisional rate (38, 39). When ions are fragmented by collisions, the rearrangement of hydrogens is very typical (40). In the solid state, reactants have contact for a prolonged time. Therefore, even the reactions that have notable barriers to proton transfer can take place due to the tunneling (24).

The formation of relatively short glycine peptide chains, observed in our experiments, provides a few possibilities for the formation of more complex molecules. The aminoketene molecule does not have a functional group that defines the type of amino acid. In our experiments, this bond is saturated with hydrogen atoms resulting in the formation of glycine residuals. However, in natural environments, other substituents could also be added to this position instead of hydrogen atoms. For example, the presence of methane or $CH_n$ radicals in the ice may allow for the formation of alanine residuals. This would likely happen by a random process leading to the formation of disordered and nonnatural peptides. However, exactly such peptides can be very interesting for the origin of life (41). There are also many efficient catalysts, including peptides themselves that can assist in the formation of peptide bonds. The presence of these catalysts may lead to the formation of much longer peptides in prebiotic conditions than those observed in the current experiments. In particular, our studies have detected the formation of Gly-Gly and Gly-Gly-Gly peptides, which have been shown to aid in the condensation of amino acid derivatives in aqueous solution (16). In this sense, the found pathway for the peptide formation is autocatalytic, which could potentially initiate natural selection in early Earth or even in the liquid phase of asteroids and comets. The peptides are also considered promising candidates for the formation of protomembranes (18). Considering all the above, the extraterrestrial peptides could have played an important role in the origin of life on Earth.

## MATERIALS AND METHODS
### Production of RTR
The experiments were performed using the same ultrahigh vacuum (UHV) setup, used in our previous studies (12). It provides a low background pressure of around $1 \times 10^{-10}$ mbar. We performed the co-deposition of gas mixtures on the surface of Si substrates cooled down to 10 K inside the UHV chamber. Together with gas mixtures, the carbon atoms generated by an atomic carbon source (42) were deposited. The source generates a pure flux of thermal carbon atoms in the ground state. The number of C atoms was at least 10 times smaller compared to the number of molecules. The fluxes of C atoms were estimated in the previous work (43). This ratio allows excluding reactions of C atoms with each other as well as with their reaction products. The chemistry on substrates was monitored by IR spectroscopy using a Fourier transform infrared spectrometer (Vertex 80v, Bruker). A quadrupole mass spectrometer (HXT300M, Hositrad) was used to monitor the gas composition in the UHV chamber. Two different types of experiments were performed. First, for better analytical characterization of RTR, we used the same (1:1, $CO/NH_3$) gas mixtures as in our previous studies and performed experiments with $^{13}C$ atoms. In the second type of experiment, defined to determine the role of water, we used $H_2O$ (49%), CO (31%), and $NH_3$ (20%) gas mixtures. For better comparison with previous results, the flux of carbon atoms was the same as in the experiments without water and we used $^{12}C$ atoms. After the deposition of the reactants, the substrate was heated at a rate of 2 K $min^{-1}$ up to 300 K, which resulted in the formation of the RTR. Alternatively, the condensation of $^{13}C$, CO, and $NH_3$ reactants was performed inside liquid helium nanodroplets (44, 45), according to the scheme depicted in fig. S6. These droplets were then collided with substrates kept at room temperature, resulting in an almost instantaneous temperature rise and evaporation of all volatiles. Therefore, this scheme provides an extremely short time during which the molecules can polymerize.

### UPLC-Q-Exactive MS analysis
Preliminary analyses were performed with UPLC coupled to an HRMS (Q-Exactive, Hybrid Quadrupole-Orbitrap Mass Spectrometer, Thermo Fisher Scientific, Waltham, MA, USA). Compound ionization was performed by heated electroSpray ionization (HESI). The separation of glycine and glycine peptides was optimized by using two hydrophilic interaction liquid chromatography (HILIC) columns in series. The first column was an Acquity UPLC Ethylene Bridged Hybrid (BEH) amide analytical column (100 mm by 2.1 mm, 1.7 μm; Waters Corporation, Milford, MA), and the second was an Acquity UPLC BEH amide analytical column (150 mm by 2.1 mm, 1.7 μm; Waters Corporation, Milford, MA).

Elution was performed at a constant flow of 200 μl $min^{-1}$ at 30°C. Acetonitrile:water (85:15) with 0.15% formic acid was used as mobile phase A and water with 0.15% formic acid and 10 mM ammonium formate as mobile phase B. The gradient started with 95% of A at 0 min was constant for 1 min and reached 80% of A in 14.80 min. A second gradient was used to reach 70% of A in 4 min and then 60% of A in 2 min. The column was then reconditioned for 2 min with 95% of A.

Mass spectrometry detection was done in full-scan mode to provide a nontargeted fingerprint of the molecules present in the sample. The electrospray voltage was set at 4.0 kV, the capillary temperature was at 280°C, and the heater temperature was at 300°C. Sheath, sweep, and auxiliary gas flow rates were set at 40, 0, and 30, respectively (arbitrary units). The Q-Exactive worked at 70,000 resolution level with an Automatic Gain Control target of 2.105, a max Ions Transfer (IT) of 150 ms, and a mass range of 70 to 1050 amu.

Exact masses of $^{12}C$- and $^{13}C$-peptide isotopes are listed in table S2. Tentative detection is provided as well.

### UPLC-TQ MS analysis
Standard solutions composed of glycine, diglycine, and triglycine (Sigma-Aldrich, Saint Louis, US) were prepared in MilliQ water. All standard solutions were diluted in methanol:water (80:20). UPLC mobile phases were prepared with LC-MS–grade acetonitrile (Sigma-Aldrich, Saint Louis, US), MS-grade formic acid (Sigma-Aldrich, Saint Louis, US), and ammonium formate salt (Fisher Chemical, Loughborough, UK).

UPLC-TQ MS analyses were performed with a Shimadzu Nexera X2 UPLC system coupled to an 8060 TQ MS (Shimadzu, Marlborough, MA, USA). Ionization was carried out by an HESI source in positive mode. The separation of glycine and glycine peptides was performed on two HILIC columns in series. The first column was an Acquity UPLC BEH amide analytical column (100 mm by 2.1 mm, 1.7 μm; Waters Corporation, Milford, MA), and the second was Acquity UPLC BEH amide analytical column (150 mm by 2.1 mm, 1.7 μm; Waters Corporation, Milford, MA). Elution was performed at a constant flow of 200 μl min⁻¹ at 30°C. Acetonitrile:water (85:15) with 0.15% formic acid was used as mobile phase A and water with 0.15% formic acid and 10 mM ammonium formate as mobile phase B. The gradient started with 95% of A at 0 min and be constant during 1 min and reached 80% of A in 14.80 min. A second gradient was used to reach 70% of A in 4 min and then 60% of A in 2 min. The column was then reconditioned for 2 min with 95% of A.

TQ MS parameters were set as follows: capillary voltage, 4 kV; interface temperature, 400°C; heat block temperature, 500°C; desolvation line temperature, 250°C; nebulizing gas flow, 3 liters min⁻¹; heating gas flow, 10 liters min⁻¹; drying gas flow, 10 liters min⁻¹. Multiple reaction monitoring (MRM) transitions of $^{12}$C and $^{13}$C-glycine and glycine peptides were verified by using a fragmentation simulation software (Mass Frontier 8.0, Thermo Fisher Scientific). MRM transitions, dwell times, and collision energies are present in table S2. We also confirmed the presence or absence of $^{13}$C molecules by comparing their elution time with the $^{12}$C isotopic standards. Since $NH_2$-Gly-$NH_2$ and $NH_2$-Gly-Gly-$NH_2$ standards were not available, we assumed that their retention times would have been similar to that of their $NH_2$-Gly$_n$-COOH counterparts, as are the retention times of $NH_2$-Gly-Gly-Gly-$NH_2$ and $NH_2$-Gly-Gly-Gly-COOH. MRM transitions, dwell times, and collision energies are present in table S2.

## Ex situ proteomic analysis
Digestion experiments were performed following the protocol described earlier (14). Briefly, the substrate was aliquoted in methanol. Ten microliters was diluted in 90 μl of phosphate buffer (pH 7). The sample was incubated at 30°C under agitation in the absence or in the presence of 0.63 μl of commercial APM [EC 3.4.11.2, aminopeptidase from porcine kidney, 3.1 mg ml⁻¹, suspension in 10 mM tris buffer salts, 3.2 mM ammonium sulfate, and 10 mM magnesium chloride (pH 7.5), Sigma-Aldrich]. The digestion was stopped by the addition of formic acid at 3.8% (v/v). To perform the analysis on the HILIC column, 20 μl of the sample was diluted in 80 μl of methanol.

## Ex situ mass spectrometry analysis
To evaluate the role of water, we used the same analytical techniques, which were used in previous work performed without water (12). Briefly, the soluble component of RTR was extracted with a water/methanol mixture (70/30) and provided for mass spectrometry analysis. It was performed using the hybrid linear trap/Orbitrap mass spectrometer (Thermo Fisher QExactive plus mass spectrometer with a heated ESI source). The accuracy of the mass determination is higher than 5 parts per million. The mass spectra were recorded after the extraction from the front side of the substrate where reactants were deposited and from the backside of the substrate, which should remain clean. The mass spectra of the material from the back side of the substrates were used to define the mass peaks that are due to the impurities that can appear at any stage of the experiment.

## Quantum chemical computations
Quantum chemical calculations were performed using the GAUSSIAN16 software (46). Geometries of the molecules and energies were determined at the B3LYP/6-311 + G (d,p) level. The reaction energies were found as the difference between the sum of the energies of reactants and the energy of the product molecules with vibrational zero-point energy corrections.

## REFERENCES AND NOTES
1. A. I. Oparin, Contemporary Theories on Origin of Life on Earth. Zhurnal Vsesoyuznogo Khimicheskogo Obshchestva Imeni D I Mendeleeva 25, 246–252 (1980).
2. M. Frenkel-Pinter, M. Samanta, G. Ashkenasy, L. J. Leman, Prebiotic peptides: Molecular hubs in the origin of life. Chem. Rev. 120, 4707–4765 (2020).
3. B. M. Rode, Peptides and the origin of life1. Peptides 20, 773–786 (1999).
4. H. D. Robertson, Life Before DNA. Science 264, 1479–1480 (1994).
5. K. J. Zahnle, R. Lupu, D. C. Catling, N. Wogan, Creation and evolution of impact-generated reduced atmospheres of early Earth. Planet. Sci. J. 1, 11 (2020).
6. B. D. Pearce, R. Molaverdikhani, R. E. Pudritz, T. Henning, K. E. Cerrillo, Toward RNA Life on early Earth: From atmospheric HCN to biomolecule production in warm little ponds. Astrophys. J. 932, 9 (2022).
7. B. K. D. Pearce, R. E. Pudritz, D. A. Semenov, T. K. Henning, Origin of the RNA world: The fate of nucleobases in warm little ponds. Proc. Natl. Acad. Sci. U.S.A. 114, 11327–11332 (2017).
8. S. Pizzarello, J. R. Cronin, Alanine enantiomers in the Murchison meteorite. Nature 394, 236–236 (1998).
9. D. P. Glavin, A. S. Burton, J. E. Elsila, J. C. Aponte, J. P. Dworkin, The search for chiral asymmetry as a potential biosignature in our solar system. Chem. Rev. 120, 4660–4689 (2020).
10. K. Altwegg, H. Balsiger, A. Bar-Nun, J.-J. Berthelier, A. Bieler, P. Bochsler, C. Briois, U. Calmonte, M. R. Combi, H. Cottin, J. De Keyser, F. Dhooghe, B. Fiethe, S. A. Fuselier, S. Gasc, T. I. Gombosi, K. C. Hansen, M. Haessig, A. Jäckel, E. Kopp, A. Korth, L. Le Roy, U. Mall, B. Marty, O. Mousis, T. Owen, H. Rème, M. Rubin, T. Sémon, C.-Y. Tzou, J. Hunter Waite, P. Wurz, Prebiotic chemicals—amino acid and phosphorus—in the coma of comet 67P/Churyumov-Gerasimenko. Sci. Adv. 2, e1600285 (2016).
11. C. Chyba, P. Thomas, L. Brookshaw, C. Sagan, Cometary delivery of organic molecules to the early Earth. Science 249, 366–373 (1990).
12. S. A. Krasnokutski, J. Chuang, C. Jäger, N. Ueberschaar, T. Henning, A pathway to peptides in space through the condensation of atomic carbon. Nat. Astron. 6, 381–386 (2022).
13. C. Eistrup, T. Henning, Chemical evolution in ices on drifting, planet-forming pebbles. Astron. Astrophys. 667, A160 (2022).
14. J. Lange, F. Djago, B. Eddhif, Q. B. Remaury, A. Ruf, N. K. V. Leitner, L. L. S. Hendecourt, G. Danger, C. G. Rodier, S. Papot, P. Poinot, A novel proteomics-based strategy for the investigation of peptide sequences in extraterrestrial samples. J. Proteome Res. 20, 1444–1450 (2020).
15. A. Warshel, P. K. Sharma, M. Kato, Y. Xiang, H. B. Liu, M. H. M. Olsson, Electrostatic basis for enzyme catalysis. Chem. Rev. 106, 3210–3235 (2006).
16. M. Gorlero, R. Wieczorek, K. Adamala, A. Giorgi, M. E. Schininà, P. Stano, P. L. Luisi, Ser-His catalyses the formation of peptides and PNAs. FEBS Lett. 583, 153–156 (2009).
17. Å. Andersson, V. Yatsyna, M. Linares, A. Rijs, V. Zhaunerchyk, Indication of 3₁₀-helix structure in gas-phase neutral pentaalanine. J. Phys. Chem. A 127, 938–945 (2023).
18. W. S. Childers, R. Ni, A. K. Mehta, D. G. Lynn, Peptide membranes in chemical evolution. Curr. Opin. Chem. Biol. 13, 652–659 (2009).
19. S. Santoso, W. Hwang, H. Hartman, S. Zhang, Self-assembly of surfactant-like peptides with variable glycine tails to form nanotubes and nanovesicles. Nano Lett. 2, 687–691 (2002).

20. C. Kocher, K. A. Dill, Origins of life: First came evolutionary dynamics. *QRB Discov.* **4**, e4 (2023).

21. N. Kitadai, S. Maruyama, Origins of building blocks of life: A review. *Geosci. Front.* **9**, 1117–1153 (2018).

22. T. D. Campbell, R. Febrian, J. T. McCarthy, H. E. Kleinschmidt, J. G. Forsythe, P. J. Bracher, Prebiotic condensation through wet–dry cycling regulated by deliquescence. *Nat. Commun.* **10**, 4508 (2019).

23. K. I. Oberg, Photochemistry and astrochemistry: Photochemical pathways to interstellar complex organic molecules. *Chem. Rev.* **116**, 9631–9663 (2016).

24. A. Potapov, S. A. Krasnokutski, C. Jäger, T. Henning, A new "non-energetic" route to complex organic molecules in astrophysical environments: The C + H₂O → H₂CO solid-state reaction. *Astrophys. J.* **920**, 111 (2021).

25. G. Molpeceres, J. Kastner, G. Fedoseev, D. Qasim, R. Schomig, H. Linnartz, T. Lamberts, Carbon atom reactivity with amorphous solid water: H₂O-catalyzed formation of H₂CO. *J. Phys. Chem. Lett.* **12**, 10854–10860 (2021).

26. S. A. Krasnokutski, C. Jäger, T. Henning, Condensation of atomic carbon: Possible routes toward glycine. *Astrophys. J.* **889**, 67 (2020).

27. S. A. Krasnokutski, Did life originate from low-temperature areas of the Universe? *Low Temp. Phys.* **47**, 199–205 (2021).

28. M. Tsuge, G. Molpeceres, Y. Aikawa, N. Watanabe, Surface diffusion of carbon atoms as a driver of interstellar organic chemistry. *Nat. Astron.* **7**, 1351–1358 (2023).

29. C. Bérubé, N. Voyer, Biomimetic epoxidation in aqueous media catalyzed by cyclic dipeptides. *Synth. Commun.* **46**, 395–403 (2016).

30. J. I. Oku, S. Inoue, Asymmetric cyanohydrin synthesis catalysed by a synthetic cyclic dipeptide. *J. Chem. Soc. Chem. Commun.* **1981**, 229–230 (1981).

31. E. R. Jarvo, G. T. Copeland, N. Papaioannou, P. J. Bonitatebus, S. J. Miller, A biomimetic approach to asymmetric acyl transfer catalysis. *J. Am. Chem. Soc.* **121**, 11638–11643 (1999).

32. W. A. Schutte, L. J. Allamandola, S. A. Sandford, Formaldehyde and organic molecule production in astrophysical ices at cryogenic temperatures. *Science* **259**, 1143–1145 (1993).

33. N. C. Wickramasinghe, Formaldehyde polymers in interstellar space. *Nature* **252**, 462–463 (1974).

34. N. P. Hammond, E. M. Parmenteir, A. C. Barr, Compaction and melt transport in ammonia-rich ice shells: Implications for the evolution of triton. *J. Geophys. Res.-Planets* **123**, 3105–3118 (2018).

35. E. L. Berger, T. J. Zega, L. P. Keller, D. S. Lauretta, Evidence for aqueous activity on comet 81P/Wild 2 from sulfide mineral assemblages in Stardust samples and CI chondrites. *Geochim. Cosmochim. Acta* **75**, 3501–3513 (2011).

36. K. T. Smith, An asteroid in the laboratory. *Science* **379**, 782–783 (2023).

37. C. M. O. Alexander, K. D. McKeegan, K. Altwegg, Water reservoirs in small planetary bodies: Meteorites, asteroids, and comets. *Space Sci. Rev.* **214**, 36 (2018).

38. R. T. Garrod, E. Herbst, Preferential destruction of NH₂-bearing complex interstellar moleculesviagas-phase proton-transfer reactions. *Faraday Discuss.* **245**, 541–568 (2023).

39. V. G. Anicich, P. Wilson, M. J. McEwan, Termolecular ion-molecule reactions in Titan's atmosphere. IV. A search made at up to 1 micron in pure hydrocarbons. *J. Am. Soc. Mass Spectrom.* **14**, 900–915 (2003).

40. D. P. Demarque, A. E. M. Crotti, R. Vessecchi, J. L. C. Lopes, N. P. Lopes, Fragmentation reactions using electrospray ionization mass spectrometry: An important tool for the structural elucidation and characterization of synthetic and natural products. *Nat. Prod. Rep.* **33**, 432–455 (2016).

41. P. Kulkarni, R. Salgia, V. N. Uversky, Intrinsic disorder, extraterrestrial peptides, and prebiotic life on the earth. *J. Biomol. Struct. Dyn.* **41**, 5481–5485 (2022).

42. S. A. Krasnokutski, F. Huisken, A simple and clean source of low-energy atomic carbon. *Appl. Phys. Lett.* **105**, 113506 (2014).

43. S. A. Krasnokutski, M. Gruenewald, C. Jäger, F. Otto, R. Forker, T. Fritz, T. Henning, Fullerene oligomers and polymers as carriers of unidentified IR emission bands. *Astrophys. J.* **874**, 149 (2019).

44. T. K. Henning, S. A. Krasnokutski, Experimental characterization of the energetics of low-temperature surface reactions. *Nat. Astron.* **3**, 568–573 (2019).

45. S. A. Krasnokutski, M. Goulart, E. B. Gordon, A. Ritsch, C. Jäger, M. Rastogi, W. Salvenmoser, T. Henning, P. Scheier, Low-temperature Condensation of Carbon. *Astrophys. J.* **847**, 89 (2017).

46. M. J. Frisch, G. W. Trucks, H. B. Schlegel, G. E. Scuseria, M. A. Robb, J. R. Cheeseman, G. Scalmani, V. Barone, G. A. Petersson, H. Nakatsuji, X. Li, M. Caricato, A. V. Marenich, J. Bloino, B. G. Janesko, R. Gomperts, B. Mennucci, H. P. Hratchian, J. V. Ortiz, A. F. Izmaylov, J. L. Sonnenberg, Williams, F. Ding, F. Lipparini, F. Egidi, J. Goings, B. Peng, A. Petrone, T. Henderson, D. Ranasinghe, V. G. Zakrzewski, J. Gao, N. Rega, G. Zheng, W. Liang, M. Hada, M. Ehara, K. Toyota, R. Fukuda, J. Hasegawa, M. Ishida, T. Nakajima, Y. Honda, O. Kitao, H. Nakai, T. Vreven, K. Throssell, J. A. Montgomery Jr., J. E. Peralta, F. Ogliaro, M. J. Bearpark, J. Heyd, E. N. Brothers, K. N. Kudin, V. N. Staroverov, T. A. Keith, R. Kobayashi, J. Normand, K. Raghavachari, A. P. Rendell, J. C. Burant, S. S. Iyengar, J. Tomasi, M. Cossi, J. M. Millam, M. Klene, C. Adamo, R. Cammi, J. W. Ochterski, R. L. Martin, K. Morokuma, O. Farkas, J. B. Foresman, D. J. Fox, Gaussian 16 Rev. C.01. (Gaussian Inc., 2016).

## Acknowledgments

**Funding:** S.A.K. acknowledges support from Deutsche Forschungsgemeinschaft DFG (grant no. KR 3995/4-2). T.H. acknowledges support from the European Research Council under the Horizon 2020 Framework program via the ERC Advanced Grant Origins 83 24 28. P.P. thanks the French National Space Agency (APR Prem's, 5472) and the French National Agency, through the PEPR Origins (ANR-22-EXOR-0014). **Author contributions:** S.A.K. conceptualized, initiated, and led the project; performed experiments with the formation of RTR, analytical characterization of RTR formed with water, data treatment, and visualization; wrote the original draft; reviewed and edited the article; and acquired financial support of the project. P.P. formulated ex situ analytical experiments and proteomics concepts, designed, and supervised these experiments, performed data treatment, acquired financial support for the project, and contributed to the writing of the original draft and data presentation. C.G. contributed to the writing of the original draft and supervised and validated ex situ analytical experiments and proteomics. Q.B.R. conducted ex situ analytical experiments and proteomics and performed data treatment. C.J. contributed to the writing of the article and the scientific discussion and provided resources. T.H. contributed to the writing of the article, the scientific discussion, and the funding acquisition. **Competing interests:** The authors declare that they have no competing interests. **Data and materials availability:** All data needed to evaluate the conclusions in the paper are present in the paper and/or the Supplementary Materials. Data that support the findings of this study are available at https://doi.org/10.7910/DVN/LED2D6.

Submitted 12 July 2023
Accepted 15 March 2024
Published 17 April 2024
10.1126/sciadv.adj7179